\documentclass[twocolumn,showpacs,amsmath,amssymb,prb,superscriptaddress]{revtex4}
\usepackage{graphicx}
\usepackage{amsmath}
\usepackage{amssymb}
\def \be {\begin{equation}}
\def \ee {\end{equation}}
\def \nn {\nonumber}
\begin{document}

\preprint{1}

\title{Self-consistent Coulomb picture of an electron-electron bilayer system}
\author{A. Siddiki}%
\address{Physics
Department, Arnold Sommerfeld center for Theoretical Physics, and
CeNS, Ludwig-Maximilians-Universit\"at M\"unchen, Theresienstr.
37, D-80333 M\"unchen, Germany}

\date{\today}
\begin{abstract}
In this work we implement the self-consistent Thomas-Fermi
approach and a local conductivity model to an electron-electron
bilayer system. The presence of an incompressible strip,
originating from screening calculations at the top (or bottom)
layer is considered as a source of an external potential
fluctuation to the bottom (or top) layer. This essentially yields
modifications to both screening properties and the
magneto-transport quantities. The effect of the temperature,
inter-layer distance and density mismatch on the density and the
potential fluctuations are investigated. It is observed that the
existence of the incompressible strips plays an important role
simply due to their poor screening properties on both screening
and the magneto-resistance(MR) properties. Here we also report and
interpret the observed MR Hysteresis within our model.
\end{abstract}

\maketitle
\section{\label{sec:1} Introduction}
The successful experimental realization of a two dimensional
electron system (2DES) has revealed a novel technique to exploit
the quantum mechanical properties of wide range of mesoscopic
systems, including integer quantized Hall\cite{vKlitzing80:494}
(IQH) and {\sl drag}\cite{Price83:750} effect. An interesting
composite two-dimensional (2D) charge system to study screening
and magneto-transport is the so called bilayer system. The basic
idea is to bring two 2DES into a close proximity, in parallel to
each other, perpendicular to the growth direction. For such a
system it was predicted, that transport in the active layer will
drive the passive layer out of equilibrium. Even if the barrier
separating the two layers is high and wide enough to prevent
tunnelling, the inter-layer interactions can still be sufficiently
strong. This effect is known as the {\sl drag
effect}\cite{Price83:750}. With the improvement of the
experimental techniques an additional electron or hole layer was
also accessible to measure magneto-transport quantities of the
bilayer systems. Motivated by the {\sl drag effect}
experiments\cite{Gramila91:1216,Sivan92:1196,Hill96:577,Rubel97:1763,
Feng97:487,Eisenstein97:486}, both the electrostatic and transport
properties of such bilayer systems were investigated theoretically
\cite{Jauho93:4420,Tso92:1333,Zheng93:8203,Halperin93:7312,Shaki97:4098,
Bonsager97:10314} within the independent electron picture,
however, the self-consistent treatment of screening was left
unresolved. On the other hand the direct measurements of the
compressibility of the 2DES\cite{Eisenstein92:674} had revealed
regimes of negative thermodynamic compressibility of the
interacting 2DES and were in qualitative agreement with the
existing theoretical predictions\cite{Tanatar89:5005}. Recently, a
magneto-resistance hysteresis has been reported for the
two-dimensional carrier systems\cite{tutuc,pan}. For the GaAs hole
bilayer system, Tutuc {\sl et al.} observed hysteretic
longitudinal resistance at the magnetic field positions where
either the majority (higher density) or minority (lower density)
layer is at Landau level filling one. They have argued that the
hysteresis is due to a layer-charge instability, which creates
domains with different layer densities, whereas for the electron
bilayer system Pan {\sl et al.} concluded that, the observed
hysteresis is due to a spontaneous charge transfer via the ohmic
contacts. The vast success of the incompressible strip
"\emph{edge}" picture explaining the IQHE rely on the
self-consistent treatment of the electron-electron interaction,
i.e. screening\cite{Lier94:7757,Oh97:13519}. One can reproduce
experimental results of the high precision \cite{Bachmair03:14}
quantized Hall (QH) plateaus and the resistance values in between
these plateaus within a relatively simple Thomas-Fermi
approximation\cite{Siddiki03:125315} and a local version of Ohm's
law\cite{Guven03:115327,siddiki2004,Siddiki04:condmat}. As a
result of self-consistent screening calculations the 2DES splits
into two "domains", namely the quasi-metallic compressible and
quasi-insulating incompressible regions. The electron distribution
within the Hall bar depends on the "pinning" of the Fermi level to
highly degenerate Landau levels. As expected, if the Fermi level
is equal to (within few $k_{B}T$, $k_{B}$ is the Boltzmann
constant and $T$ is temperature) a Landau level with high density
of states (DOS) the electron system is known to be compressible
(locally), other wise incompressible. In a very recent experiment
similar magneto-transport hysteresis was
observed\cite{siddikikraus:05} at an electron-electron bilayer
system. It was attributed to thermodynamical non-equilibrium
caused by the incompressible strips similar to the explanation of
the hysteresis also observed at single layer
systems\cite{jhuels2004}. The experimental findings were
interpreted as the fingerprints of dissipationless eddy currents
driven by an induced gradient in the electrochemical potential
over the incompressible regions. Huels \emph{et al.} concluded
that, when measuring the Hall resistance while sweeping the
magnetic field, the 2DES cannot be considered as being at
thermodynamic equilibrium at the plateau regimes, where one
expects incompressible regions.

In this paper we apply the self-consistent scheme developed in the
previous
works\cite{Siddiki03:125315,Guven03:115327,siddiki2004,Siddiki:ijmp}
to an electron-electron bilayer system. For this model system, we
can investigate both self-consistent screening and
magneto-transport properties within the linear response regime
under QH conditions. We show that the existence of the
incompressible regions in one of the layers, affects the other
layer density profile strongly by creating potential fluctuations.
We observe that these potential fluctuations modify the
magneto-transport quantities. Here we present our model in detail,
that explains some of the recent experimental
findings\cite{siddikikraus:05}, where the magneto-transport
hysteresis was observed for mismatched densities, whenever one of
the layers is in the QH plateau regime. 

The organization of this work is as follows: first the bilayer
geometry and fixed external potentials acting on the electron
layers are introduced. Then we discuss the electron density and
electrostatic potential profiles of the system, which are obtained
within the self-consistent Thomas-Fermi-Poisson
approach\cite{Siddiki03:125315}. Second, we systematically
investigate the influence of the incompressible strips depending
on temperature, inter-layer distance and the density mismatch,
then shortly discuss the effect of density mismatch on both
electrostatic and transport properties. Here, we use the scheme
presented by Siddiki and Gerhardts (SG)\cite{siddiki2004}, to
calculate the Hall and the longitudinal resistances. We examine
the effects of the incompressible strips from the point of
external potential fluctuations based on the arguments given in
the same work. There they have concluded that a perturbing
external potential may shift, widen and/or stabilize the QH
plateaus.
\section{\label{sec:2} The Model}
In a typical electron bilayer sample, a silicon doped thick
$\rm{(AlGa)As}$ layer is grown on top of a $\rm{GaAs}$ substrate.
This is followed by the bottom $\rm{GaAs}$ quantum well, separated
from the top 2DES by an un-doped $\rm{AlGaAs}$ spacer. On top of
the upper 2DES again a silicon doped $\rm{(AlGa)As}$ layer is
grown, which is capped by a $\rm{GaAs}$ layer. So that two 2DES
are placed in a close proximity, which are confined by remote
donors
 and the sample is capped by top and/or bottom gates that controls
the electron density of each layer. In related experiments both
gates are used to tune the electron densities from top and bottom
layers, whereas the gate potential profile depends on the sample
geometry and the applied gate bias. The electrons are
symmetrically (with respect to the growth direction) confined by
$\rm{AlGaAs}$ layers each of which contains a plain with $\rm{Si}$
doping ($\delta-$ doping), and are separated by a spacer of
thickness ($h\sim10-30$nm). For such a separation thickness the
bilayer system is then known to be electronically decoupled and in
non-equilibrium, i.e. can be represented by two different
electrochemical potentials. Electron tunnelling between the layers
is not possible.

\begin{figure}[h]
{\centering
\includegraphics[width=.7\linewidth]{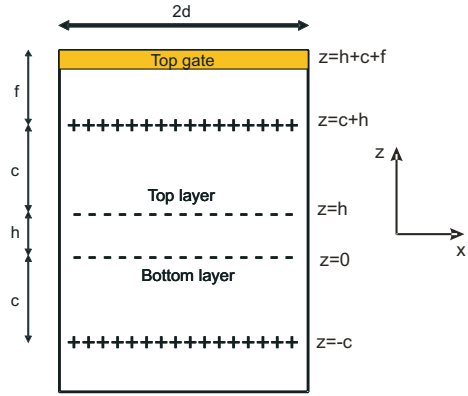}
%
\caption{ \label{fig:geometry} The cross-section of
the bilayer geometry. Sample is capped by a gate from top in order
to change the electron density of the top-layer. Top (bottom) 2DES
lies below (above) the top (bottom) layer donors at a distance,
$c$. Electron layers are separated from each other by a dielectric
spacer having thickness, $h$.}}
\end{figure}

In Fig.~\ref{fig:geometry} a schematic drawing of a mesa etched
bilayer system is shown, which consists of two 2DESs (minus
signs), two donor layers (plus signs) and a top gate (gray area).
We model the bilayer system such that the bottom 2DES lies on the
$z=0$ plane with a number density $n^{B}_{\rm el}(x)$ in the
stripe $-d<x<d$, where $2d$ is the width of the sample and the top
layer is in the plane $z=h$, with the electron density $n^{T}_{\rm
el}(x)$. Relevant to the considered experiments, we assume that
the donors (or equivalently the background charges) are
symmetrically separated from the electron layers placed at $z=-c$
and $z=c+h$, for bottom and top layers respectively, having a
constant surface density $n^{T}_{0}=n^{B}_{0}=n_{0}$. Here we also
assume translation invariance in the $y$ direction. The electron
density of the top 2DES is governed by the top gate, located at
$z=c+h+f=z_{g}$ and the electron channels are formed in the
interval $|x|\leq b$.
One obtains the total electrostatic potential of an electron on
the line $(x,z)$ due to a line-charge at $(x_{0},z_{0})$
from\cite{Siddiki03:125315} \be V(x,z)= V_{bg}(x,z) + V_H(x,z),
\label{eq:Vvon_x}\ee and \be V_H(x,z)= \frac{2e^2}{\bar{\kappa}}
\int_{x_{l}}^{x_{r}} dx_{0} K(x,x_{0},z,z_{0}) n_{\rm
el}(x_{0},z_{0}), \label{eq:V_Hartree} \ee where $-e$ is the
charge of an electron, $\bar{\kappa}$ an average background
dielectric constant, and the kernel $K(x,x_{0},z,z_{0})$ solves
Poisson's equation under the relevant boundary conditions given by
\be \label{perp-bilayer} K(x,x_{0},z,z_{0})=-\frac{1}{2}\ln\!
\left(\frac{\cos ^2 \frac{\pi}{4d}(x+x_{0}) + \gamma ^2}{ \sin ^2
\frac{\pi}{4d} (x-x_{0}) + \gamma ^2}\right) \ee where the $z-$
dependence is given by $\gamma=\sinh (\pi |z-z_{0}|/4d)$, due to a
missprint a factor $1/2$ was missing in our previous work,
although the numerics included this factor. The confining
(background) potential is obtained by inserting a constant number
density ($n_{0}$) of the background charges into
Eq.~(\ref{eq:V_Hartree}). It is assumed that the gate can be
described by an induced charge distribution $n_{g}(x)$, residing
on the plane $z_{g}$. Then, from Eq.(\ref{eq:V_Hartree}) the gate
potential can be written as \be
V_{g}(x,z)=\frac{2e^2}{\bar{\kappa}} \int_{-d}^{d} dx_{0}
K(x,x_{0},z,z_{g})n^{}_{\rm g}(x_{0}). \label{eq:gate} \ee
\begin{figure}[h]
{\centering
  \includegraphics[width=1.\linewidth,angle=0]{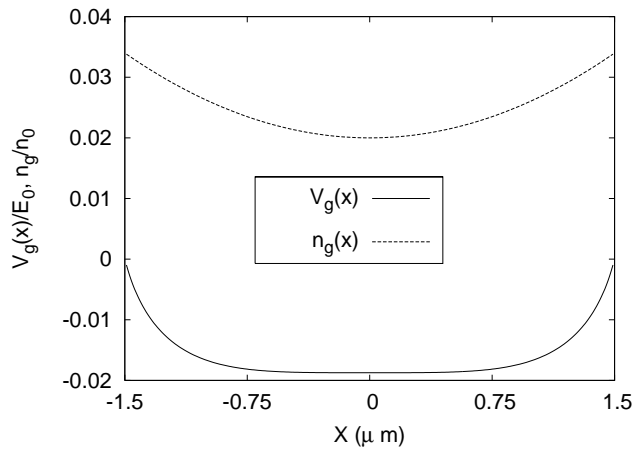}
\caption{ \label{fig:gatepotential} The gate potential profile
(solid-line) at $z=h$, together with the generating positive
charge distribution (dotted-line) against position. The additional
charges residing on the gate lead to a stronger confining
potential in the top layer, resulting in a higher average electron
density. The gate potential strength is taken to be
$n_{g}^{0}/n_{0}=0.02$, where $E_{0}$ is the pinch-off energy. It
is defined as $E_{0}=(2\pi e^2 n_0 d/\bar{\kappa})(\pi^{2}/8G)$
normalized with $G=0.915965594$, the Catalan's
constant\cite{Gradshteyn} for $\gamma=0$. The donor density is
$n_{0}=4\cdot10^{11}cm^{-2}$. For other parameters see text.
 }}
\end{figure}
In order to obtain a flat (gate) potential profile at the bulk we
choose the induced charge distribution as \be
n_{g}(x)=n_{g}^{0}(1+\alpha(x/d)^2) \label{gate},\ee where
$n_{g}^{0}$ determines the strength of the gate potential whereas
$\alpha$ gives the slope of the induced charge distribution. In
Fig.~\ref{fig:gatepotential} the distribution of the positive
charges and the resulting potential profile is shown for
$\alpha=0.7$. One can add more electrons to the top layer by
setting $n_{g}^{0}/n_{0}$ to a positive value, while keeping the
depletion length $d-b$ fixed at zero temperature and for vanishing
magnetic field. With such treatment of the gate, the average
electron densities are changed while the depletion length is kept
constant. Similarly the Hartree potential is calculated from
Eq.(\ref{eq:V_Hartree}) for the top layer as \be
V_{H}^{T}(x,z)=\frac{2e^{2}}{\bar{\kappa}} \int_{-d}^{d} \!dx_{0}
K(x,x_{0},z,z_{0}=h)n^{T}_{\rm el}(x_{0}) \nn \ee and for the
bottom layer \be V_{H}^{B}(x,z)=\frac{2e^{2}}{\bar{\kappa}}
\int_{-d}^{d} \!dx_{0} K(x,x_{0},z,z_{0}=0)n^{B}_{\rm el}(x_{0}) .
\nn \ee Then Eq.(\ref{eq:Vvon_x}) can be rewritten as \be V(x,z)=
-V^{T}_{bg}(x,z)-V^{B}_{bg}(x,z)-V_{g}(x,z)\nn \ee \be
+V_{H}^{T}(x,z)+V_{H}^{B}(x,z) .\label{int1d}\ee In the next step
the electron densities are calculated within the TFA: \be
\label{eq:TFA} n^{T,B}_{el}(x)=\int
dED(E)f([E+V(x,z_{T,B})-\mu^{*}_{T,B}]/k_{B}T), \ee with $D(E)$
the Landau density of states (DOS), $f(E)=1/[\exp(E/k_BT)+1]$ the
Fermi function, $\mu^{*}_{T,B}$ is the chemical potential (being
constant in the equilibrium state) and $z_{T,B}$ the position of
top and bottom layer, respectively. We will use the Landau DOS
described by
\be \label{landau-dos} D(E)=\frac{1}{\pi
l^{2}}\sum_{n=0}^{\infty}{\delta (E-E_n)},
 \ee
as default, unless other definitions are given. In our numerical
calculations, we start with zero temperature and magnetic field
and initially obtain the confining potential created by its own
donors for each layer.
\begin{figure}[h]
{\centering
  \includegraphics[width=.8\linewidth,angle=-90]{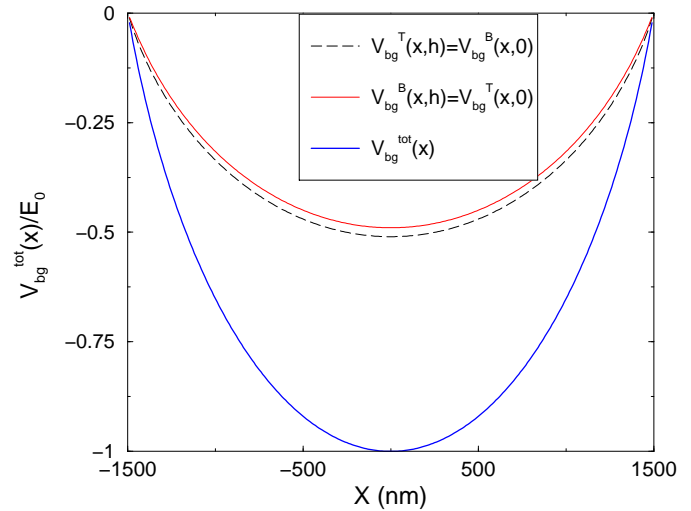}%
%
\caption{\label{fig:background}Total background potential created
by the top and bottom layer donors (thick solid-line). The
potential contributions to the bottom 2DES $(z=0)$ come from top
donor layer ($V_{bg}^{T}(x,0)$, thin solid line) and from bottom
donor layer ($V_{bg}^{B}(x,0)$, dashed line). Numerical
calculations are performed for GaAs heterostructure with the
parameters $h=15$nm and $c=60nm$. The numbers of donors are fixed
to $4\times10^{11}$cm$^{-2}$ per layer, for a sample width
$2d\sim3\mu$nm, with  an average pinch-off energy $E_{0}$.
 }}
\end{figure}
Then, we obtain the electron densities and the Hartree potentials
of both layers just as in the case of a single
layer\cite{Siddiki03:125315}.
 Knowing the electron distribution
of the top (bottom) layer via Eqs.(\ref{eq:TFA}) and
(\ref{eq:V_Hartree}) we calculate the potential acting on the
bottom (top) layer similar to the single layer case. So that the
total electrostatic potential of an electron at the top layer is
\be V^{T}(x)=V_{ext}(x,h)+V^{B}_{H}(x,h)+V^{T}_{H}(x,h),
\label{vt}\ee where the total fixed external potential acting on
the layers is $V_{ext}(x,z)=V^{tot}_{bg}(x,z)+V_{g}(x,z)$ and we
denote the total background potential by
$V^{tot}_{bg}(x,z)=V^{T}_{bg}(x,z)+V^{B}_{bg}(x,z)$ (see
Fig.\ref{fig:background}). Note that the first two terms in
Eq.~(\ref{vt}), are considered to be external and the last term is
the intra-layer Hartree potential. Similarly for the bottom layer
the total electrostatic potential is \be V^{B}(x)=V_{ext}(x,0)
 +V^{T}_{H}(x,0) +V^{B}_{H}(x,0). \ee

In each iteration step an accurate numerical convergency is
achieved for a single layer, then the Hartree potential is added
to the other layers external potential. Intra-layer
self-consistencies are obtained by Newton-Raphson iteration and
inter-layer self-consistency is obtained by direct iteration. One
can then use this solution as an initial value and obtain the
density and screened potential for finite magnetic field and
temperature. We scale energies by the average pinch-off energy
$E^{*}_{0}=(E^{T}_{0}+E^{B}_{0})/2=E_0$, since we always assume
symmetric donor distribution, e.g.
$\Omega_{c}=\hbar\omega_{c}/E_{0}$ (and set the gate voltage to be
zero, thus no extra charges reside on the top gate). The lengths
are scaled by the screening length $a_{0}$ ($=a^{*}_{B}/2$)
expressed in terms of effective Bohr radius,
$a^{*}_{B}=\bar{\kappa}\hbar^{2}/(me^{2})$. The electron density
and the electrostatic potential can be calculated
self-consistently by the above scheme within the TFA.
\section{Screening results}
In this section we present our results calculated within the
self-consistent scheme, starting by discussing the zero
temperature, zero magnetic field limit and then compare the
obtained density profiles with the finite temperature and field
profiles. The aim of this investigation is to clarify the effect
of the quantizing perpendicular magnetic field, which introduces
local charge imbalances due to formation of the incompressible
strips. We have already seen that the electron density
distribution is highly sensitive to the applied external magnetic
and electric fields. Therefore even very small changes in these
external parameters affect the density and potential profiles
drastically.

\begin{figure}[t] {\centering
\includegraphics[width=.5\linewidth,angle=0]{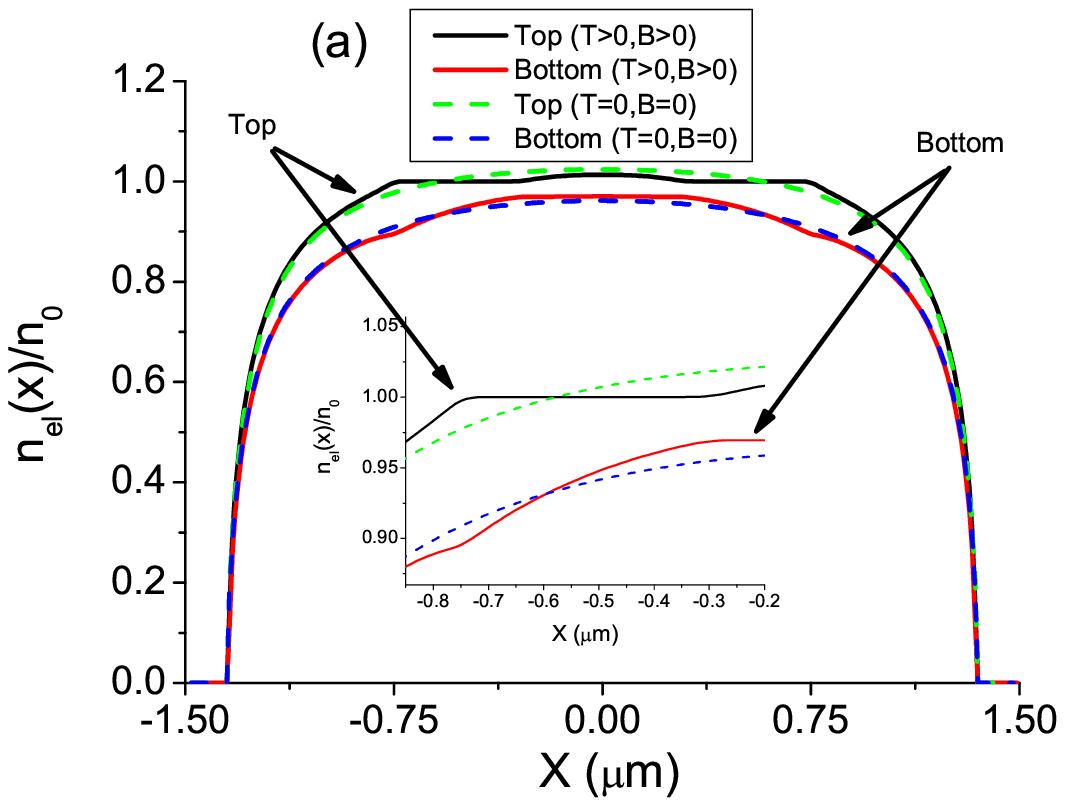}
\includegraphics[width=.5\linewidth,angle=0]{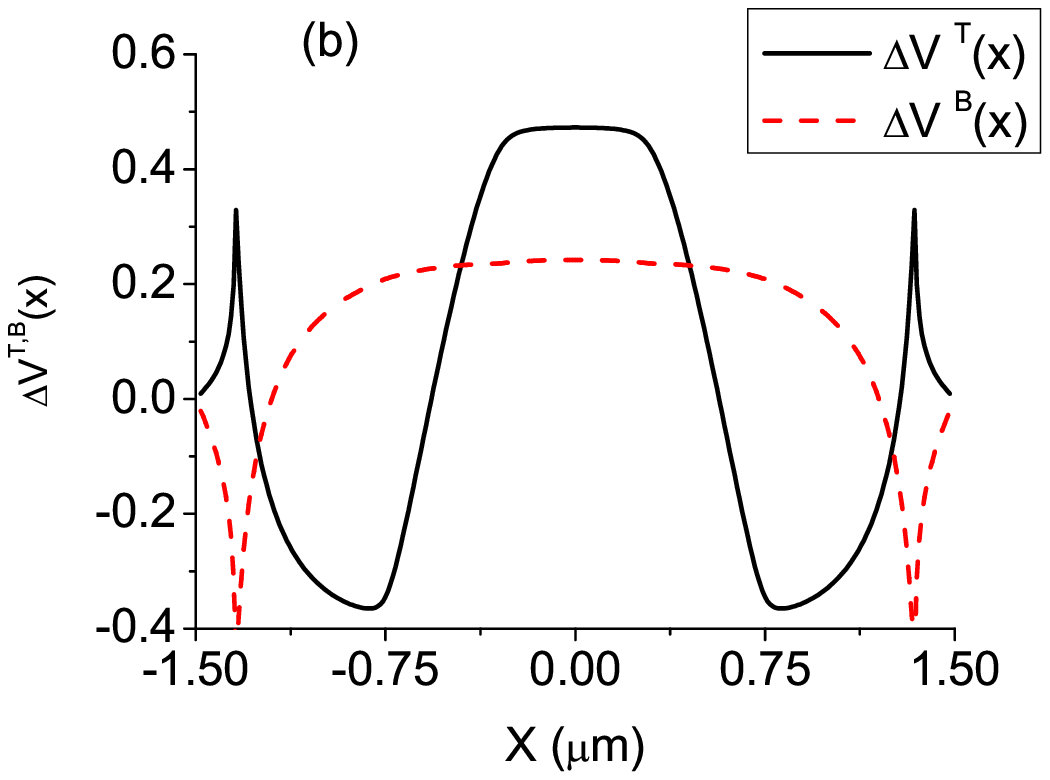}
\caption{\label{fig:tbzero}[a] Electron densities for a finite
magnetic field (such that, $\bar{\nu}^{tot}_{av}=1.6$) and at
default temperature ($kT/E_{0}=5\times10^{-5}$ or
$kT/E_{F}\sim0.01$), for top and bottom (thin solid-line) layers.
Also for vanishing field and temperature (dotted lines). [b] The
resulting potential variation due to the formation of the
incompressible strips, where the superscripts refers to top (T)
and bottom (B). The depletion length is set to be $150$nm and the
density mismatch is governed by applying a finite gate potential
$V_{0}/E_{0}^{*}=0.02$, resulting in
$\bar{n}_{el}^{T}=3.42\times10^{11}$cm$^{-2}$ and
$\bar{n}_{el}^{B}=3.18\times10^{11}$cm$^{-2}$. The inset shows the
region where an incompressible strip exists at the top layer.}}
\end{figure}

It is well known\cite{Siddiki03:125315} that the formation of the
compressible and incompressible strips results in an inhomogeneous
density distribution that deviates from the zero field profile.
This deviation creates a local charge imbalance generating a
potential fluctuation. We describe this fluctuation by, \be \Delta
V(x)=[V(x,0,0)-V(x,T,B)]/ \Omega_{2}, \ee where the
self-consistent potentials are calculated at zero and finite
magnetic field and temperature, respectively (see
Fig.\ref{fig:tbzero}b). Here after $\Omega_{2}$
($=\hbar\omega_{c}/E_{0}=0.328\times10^{-2}$) represents the
dimensionless cyclotron energy at average filling factor two,
since we always consider situations, where average filling factor
is around two. We start our analysis by discussing the effect of
the local charge deviation from its equilibrium distribution at
$T=0\,,B=0$. Figure ~\ref{fig:tbzero}a depicts the electron
densities calculated within the SCTFPA for vanishing and finite
magnetic field and temperature, where we used $300$ mesh points to
span a single layer. The typical parameters used in the
calculations are: $d=1570$nm, $c=60$nm, $h=15$nm, $f=100$nm,
$n_{0}=4\times10^{11}$cm$^{-2}$ and for unbiased gate
$\bar{n}^{T}_{el}=\bar{n}^{B}_{el}=3.31\times10^{11}$cm$^{-2}$
which result in a Fermi energy ($E_F$) $\sim14$meV.
 A positive (with respect to the electrons)
potential bias is applied to the top gate [for details check
Eq.(\ref{gate}) and related text] so that more electrons are
populated to the top layer resulting in a density mismatch. The
curves for finite field and temperature show a considerable
deviation from the curves for zero field and temperature in the
intervals ($200$nm$<|x|<1000$nm), where one observes
incompressible strips at the top layer. In the inset we
concentrate on this interval. In Fig.~\ref{fig:tbzero}a we compare
the $T=0,B=0$ density curves to the $T\neq0,B\neq0$ ones. In the
interval $-600$nm$<x\lesssim -300$nm there are less electrons at
the top layer due to the formation of an incompressible strip.
This yields a less repulsive inter-layer Coulomb interaction. So
that more electrons are populated locally at the bottom layer.
Similar arguments hold for the left-hand side of the
incompressible strip ($-800$nm$\lesssim x<-600$nm) but now the
potential becomes more repulsive thus more electrons are depleted
from the bottom layer locally. The corresponding potential
variations are shown in Fig.\ref{fig:tbzero}b, the peak like
behavior near the edges results from temperature difference,
whereas the structure observed at the top layer is due to the
formation of the incompressible strip in the same layer. Since the
bottom layer is completely compressible the potential variation
does not show any non-monotonous feature and the fluctuation is
perfectly screened. A relatively large variation is seen within
the depletion regions. The reason for this is that, at finite
temperature the density profiles leak out at the edges. At very
strong magnetic fields ($\nu(0)<2$) only the lowest Landau level
is partially occupied, i.e. high DOS, thus the density profiles
for vanishing and for very strong magnetic field should look very
similar. However, for intermediate $B$ there is a difference due
to the quantizing magnetic field, that creates dipolar
(incompressible) strips at the top layer, which have an influence
on the bottom layer via the strong Coulomb interaction. Therefore,
it is essential to examine the formation of the incompressible
strips at strong magnetic fields. This is done by manipulating the
sheet electron densities by applying a finite gate bias. The gate
controls the existence and positions of the incompressible strips
indirectly, so that one can examine the screening effects of the
incompressible strips considering the electron density
distributions. In the later sections, potential fluctuations
created by these local charge imbalances, i.e. the incompressible
strips, will be connected to the magneto-transport quantities,
where we explicitly show the impact of the incompressible strips
on the Hall resistances.

\subsection{Intra-layer distance, temperature and density mismatch}
\begin{figure}[h] {\centering
  \includegraphics[width=1.\linewidth,angle=0]{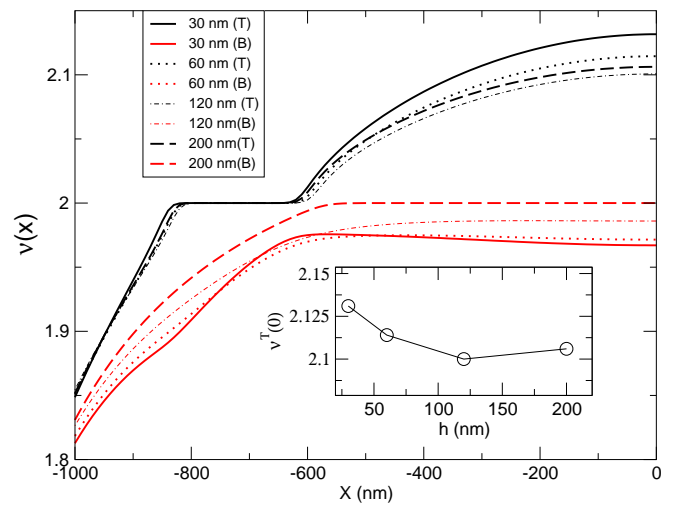}
\caption{\label{fig:intradistance}Local filling factors versus
position at a finite magnetic field and for $h=30,60,120$ and
$200$ nm, at default temperature. The top layer (upper set)
exhibits an incompressible strip and whereas bottom layer (lower
set) is compressible all over the electron channel, except
$h=200$nm. The inset depicts the center filling factor of the top
layer for four $h$ values. }}
\end{figure}

It is well known that the mutual Coulomb interaction is a strong
long-range interaction. Thus, a change in the charge distribution,
compared to the equilibrium distribution (at $T=0\,,B=0$),
produces a considerable effect on the observable quantities even
at large distances. In the previous section it is shown that such
a local charge imbalance, connected with a potential fluctuation,
is created due to formation of the incompressible strips. Here we
investigate the density distributions of the layers by applying a
positive gate voltage, hence populating the top 2DES and vary the
intra-layer distance and temperature. The average total filling
factor, $\bar{\nu}^{tot}_{av}=(\nu^{T}+\nu^{B})/2$, is kept
constant and the evolution of the incompressible strip at the top
layer, and its effect to the bottom layer, is examined. In
Fig.~\ref{fig:intradistance} we show the local filling factors of
both layers versus position, where the top layer (upper set of
curves) has an incompressible strip in the interval
$-850$nm$<x<-600$nm for different inter-layer distances. The
influence of the incompressible strip in the top layer on the
electron distribution of the bottom layer disappears rapidly
although the intra-layer distance is changed rather smoothly. This
is due to exponential decay of the amplitude of the Coulomb
potential in the $z-$ direction, i.e. $V(q,z)\sim
V_{q}\exp{-q|z|}$. The effective confining potential (ECP)
experienced by each layer, of course, depends strongly on the
inter-layer distance, hence for each $h$ value the number of
electrons at each layer changes. This is depicted at the inset of
Fig.\ref{fig:intradistance}, there we concentrate to the center
filling factor center. We observe that by increasing $h$($=30,60$
and $120$nm) the number of electrons at the center decreases (at
fixed magnetic field) and the ECP becomes less confining thus a
flatter density distribution is observed. Interestingly, for the
largest separation ($h=200$nm) the bottom layer becomes widely
incompressible at the bulk, hence semi-transparent, and the
electrons of the top layer start to \textit{see} the donors of the
bottom layer. This specific configuration leads accumulation of
the top layer electrons to the bulk and is a clear indication of
non-linear screening, due to incompressible strips, in such a
bilayer system. This fact emphasizes that even small a change at
$h$ has a strong influence on the electron density profile. The
exponential decay of the Coulomb interaction in the growth
direction essentially determines whether the two 2DESs are
strongly coupled or not, depending on the formation of the
incompressible strips, i.e. screening. In the relevant experiments
the distance between the two layers is fixed during the growth
process. Hence we proceed our investigation by fixing $h=15$nm and
vary the electron temperature. For such a separation the system is
known to be electronically uncoupled and can be described by two
electrochemical potentials.

\begin{figure}[h] {\centering
  \includegraphics[width=1.\linewidth,angle=0]{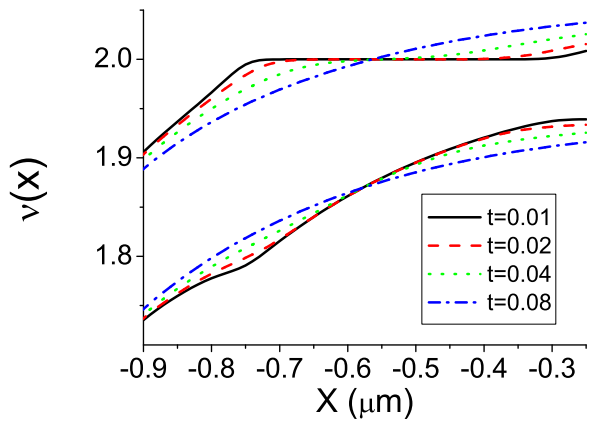}
\caption{\label{fig:tcompare} Density profiles across the
incompressible strip at the top layer as a function of
temperature. For a fixed density mismatch,
$\bar{n}_{el}^{B}/\bar{n}_{el}^{T}=0.93$, at
$\bar{\nu}^{tot}_{av}=1.6$ with $h=15$nm.}}
\end{figure}

In Fig.~\ref{fig:tcompare} the temperature dependence of the
incompressible strip residing in the top layer and the
compressible region at the bottom layer is shown. At single layer
geometries it is well known\cite{Lier94:7757,Oh97:13519} that the
quantizing effect of the magnetic field becomes inefficient, if
the thermal energy of the system becomes larger than few percents
of the cyclotron energy. Here we also observe a similar behavior
for the bilayer system where the incompressible strip at the top
layer and, moreover, the local charge imbalance seen at the bottom
layer disappears while increasing the dimensionless temperature
$t(=kT/E_F)$ from 0.01 to 0.08. This fact shows that the local
inhomogeneity at the bottom layer density distribution is only due
to the dipolar strip at the top layer and is very sensitive to the
temperature and exists for $t<0.04$.

\begin{figure}[h]
{\centering
  \includegraphics[width=1.\linewidth,angle=0]{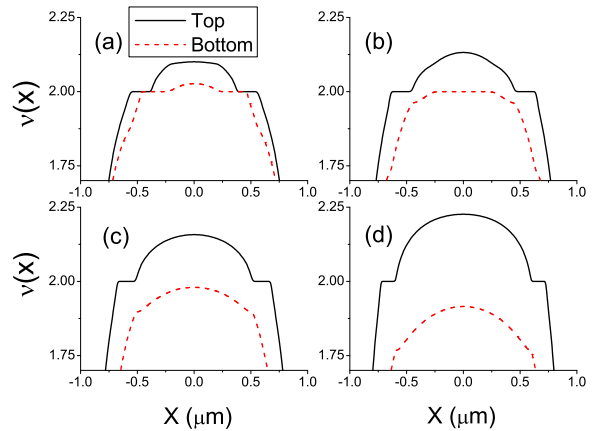}
\caption{ \label{fig:mismatch} The top layer is populated by
applying a finite gate potential $V_{0}/E_{0}^{*}$ [a] 0.01, [b]
0.02, [c] 0.03, [d] 0.05 corresponding to density mismatches of
$\bar{n}_{el}^{B}/\bar{n}_{el}^{T}=0.96, 0.93, 0.89, 0.84$,
respectively. Given at default temperature, with the depletion
length, $b/d=0.9$ and $\bar{\nu}_{av}^{tot}=1.67$.
 }}
\end{figure}

Relevant to the experiments considered, density mismatch is
another tunable parameter, like the temperature. For this purpose
now we fix the temperature and intra-layer distance and vary the
potential of the top gate essentially by changing the number of
induced positive charges. Figure \ref{fig:mismatch} presents
position dependent filling factors for such a positively charged
gate, simulating different density mismatches by applying a gate
bias voltage $V_{0}$. For a slight mismatch one does not observe a
prominent change, except more electrons are accumulated at the
center of the top layer. Due to the stronger Coulomb repulsion the
bulk of the bottom layer is more depopulated and shows a flatter
profile ( see Fig.\ref{fig:mismatch}a). In
Fig.\ref{fig:mismatch}b, this feature is more pronounced at a
higher gate bias and, in turn, the bottom layer is forced to be
incompressible at the bulk, leading the effective external
potential to be more confining for the top layer. The outer edge
regions of incompressible strips residing at the top layer
suppress the electrons beneath, at the bottom layer. Increasing
the density mismatch in favor of the top layer, in
Fig.\ref{fig:mismatch}c and d, it is observed that more electrons
start to accumulate at the bulk of the top layer and small density
fluctuations can be seen at the bottom layer, however, we do not
see any prominent change at the density profiles since the bottom
layer can screen perfectly. We close our short discussion by
noting that, even a small amount of density mismatch can lead to a
drastic change in the density profiles. Moreover this change is
enhanced by the existence of the incompressible strip, which
results from the non linear screening in the presence of external
magnetic field. This high sensitivity to the external electric
field (here the gate) is clearly seen, if the potential profiles
are considered. We shall emphasize, that a wide incompressible
strip formed at the bulk of one of the layers does not necessarily
imply that this layer becomes completely transparent to its
donors, since the electrons residing in the incompressible strip
create an electric field which can still partially cancel the
electric field generated by its donors.

\subsection{\label{sec:potentialfluctuations}Potential fluctuations}
\begin{figure}[h]
{\centering
  \includegraphics[width=1.\linewidth,angle=0]{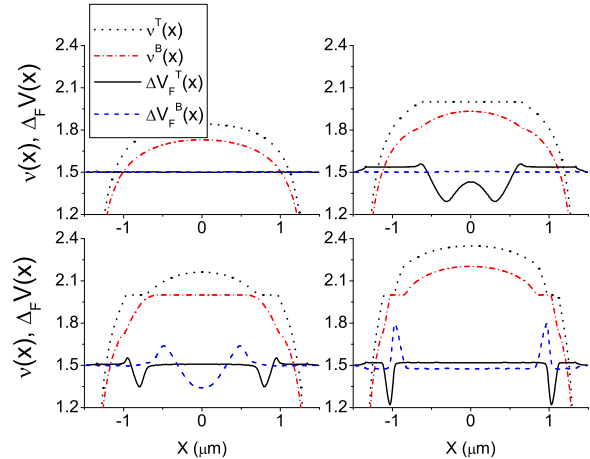}
\caption{ \label{fig:potfluc} Position dependent filling factors
and magnetic field induced potential variations, for
$\bar{\nu}_{av}^{tot}$ [a] 1.44 [b] 1.59 [c] 1.69 [d] 1.84. For
clarity all the potential variations are shifted by an amount of
1.5. The gate potential is set to $V_{0}/E_{0}^{*}=0.02$, for
fixed depletion and at default temperature.
 }}
\end{figure}

In this section we examine the effects of the incompressible
strips in one layer on the external potential profile of the other
layer by comparing what we call "interacting" and
"non-interacting" systems.
 In both cases we start with the self-consistent calculation of the density
 and potential profiles at $T=0$, $B=0$. Then we focus on one layer,
 bottom or top, which we call the "active" layer while the other layer is
 called the "passive" one. Now we keep the $T=0$, $B=0$ density and Hartree
 potential profile of the passive layer fixed and calculate in the corresponding,
 $B-$ independent external potential of the active layer its density and potential
 profile self-consistently at finite $T$ and $B$. This yields the total potential
 $V_{N}(x,z)$ of the non-interacting system. It takes into account the $B-$ dependent
 intra-layer screening  properties in the active layer, but not the $B-$ dependent
 changes of the intra-layer screening in the passive layer and of the inter-layer
 screening. Finally we drop the restriction on the passive layer and calculate
 density and potential profiles for both layers at finite $T$ and $B$ fully self-consistently.
 This yields the total potential $V_{I}(x,z)$ of the interacting system.
\newline The potential variation (at finite temperature and
magnetic field, indexed by $F$ as a subscript) \be \Delta_{F}
V(x)=[V_{I}(x,z)-V_{N}(x,z)]/\Omega_{2}, \ee taken at the $z-$
value of the active layer, describes the $B-$ dependent change of
the interaction of the active with the passive layer. It is
suitable for studying the effect of the incompressible strips in
the passive layer on the effective potential in the active layer.
 Fig.~\ref{fig:potfluc} shows potential variation
and filling factors across the sample at four different magnetic
field values where the superscript $T(B)$ indicates that the top
(bottom) layer is the active layer. In Fig.~\ref{fig:potfluc}a the
magnetic field strength is chosen such that both layers are
compressible, i.e. the center filling factors of both layers are
slightly below two. The potential variation shows a characteristic
behavior as the layers are both compressible all over the sample,
the screening is nearly perfect and we do not observe any
significant potential fluctuation. If one decreases the magnetic
field strength and obtains a wide incompressible strip in the bulk
of the top layer (Fig.~\ref{fig:potfluc}b), a large potential
variation is observed at this layer. This is nothing but the
charge quadrupole at the center, treated by
Ref.[\onlinecite{Chklovskii93:12605}] for a single layer geometry.
The variation drastically increases and becomes almost $\sim \%30
\Omega_{2}$. Meanwhile, the variation of the bottom layer does not
show any significant change. The explanation of these observations
is twofold; first as an incompressible strip is formed at the top
layer, the finite field density profile strongly deviates from the
zero field profile, which essentially creates a huge local (within
the incompressible strip) external potential fluctuation to the
bottom layer. Second, the bottom layer is completely compressible,
so that this large potential fluctuation can be screened by
redistribution of the bottom layer electrons, resulting in a
density deviation from the zero $B$ profile. Note that as a
consequence of the self-consistency the deviation is spread all
over the bottom layer, which also generates an external potential
fluctuation to the top layer. Here, as we examine how the top
layer responds to this external potential fluctuation, we remind
the reader that there are two different regions with different
screening properties: (i) The compressible regions
($0.5\mu$m$<|x|<1.5\mu$m interval in Fig.\ref{fig:potfluc}b) show
similar features to the bottom layer and the external potential
fluctuation is well screened, (ii) The incompressible region
($0\mu$m$<|x|<0.5\mu$m interval in Fig.\ref{fig:potfluc}b) at the
bulk can poorly screen the fluctuation and one observes that
$\Delta_{F} V(x)$ can be large as half the Fermi energy.
In Fig.~\ref{fig:potfluc}c, the magnetic field strength is
slightly decreased in order to obtain a wide incompressible strip
at the bottom layer, meanwhile the wide incompressible strip of
the top layer splits into two ribbons which are shifted towards
the edges. For this $B$ value, one observes both the quadrupole
(for the bottom layer) and the dipole (at the top layer) moments.
Figure ~\ref{fig:potfluc}d shows the case, where two well
developed incompressible strips are present at both of the layers
and the potential fluctuations, created by the charge dipoles, are
confined to these regions. Different than the previous case we
only observe one side of the dipole moment, since the other half
is screened by the other layer. We should emphasize that the
variation amplitude depends on the width of the incompressible
strip which generates the fluctuation, since the charge imbalance
becomes larger if the incompressible strip is wide. The finding is
simply that the potential fluctuations in a layer exist only
within incompressible regions and are screened in compressible
ones.

\subsection{Modulated donor distribution}
It is rather a common technique, to add an external (harmonic)
modulation potential to the confinement potential in order to
study the peculiar low-temperature screening properties of a 2DES,
which has been done for a single layer geometry
\cite{Wulf88:4218,Siddiki02:Oxford,Siddiki03:125315}. Although a
double layer geometry is a very promising system, even a
qualitative investigation of such a modulation is still missing in
the literature. In this section we provide a simple model to
discuss the effect of this modulation on screening and compare our
results, qualitatively, with the single layer ones.

\begin{figure}[h]
{\centering
  \includegraphics[width=1.\linewidth,angle=0]{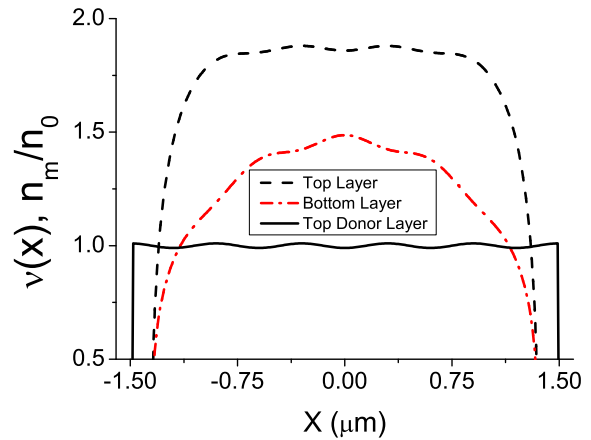}
\caption{ \label{fig:donor} Spatial distribution of the donors
(solid line), together with the local filling factors of top
(dashes) and bottom (dashed-dots) layers at default temperature.}}
\end{figure}

\begin{figure}[h]
{\centering
  \includegraphics[width=1.\linewidth,angle=0]{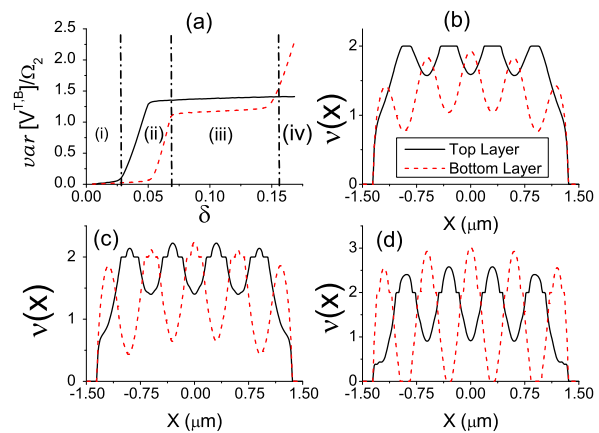}
\caption{ \label{fig:mod} [a] The variation of the screened
potential versus the modulation amplitude, together with the
typical electron densities corresponding to different screening
regimes  [b] ii, $\delta=0.026$, [c] iii, $\delta=0.073$ and [d]
iv, $\delta=0.157$. For top (solid lines) and bottom (dashed
lines) layers.}}
\end{figure}

A harmonic density modulation is added to the background charge
distribution in order to obtain a potential modulation. Here we
should also note that such a modulation will be used to simulate
the long-range fluctuations of the confining potential in the
following sections\cite{Siddiki:ijmp}. It is assumed that the
spatial distribution of the donors is given by \be
n_{m}(x)=n_{0}\big(1-\delta\cos(\lambda\pi (x/2d))\big)
\label{donormod}\ee where $\lambda$ is an odd integer preserving
the boundary conditions and $\delta$ describes the strength of the
modulation. In the following discussion we will fix the density
mismatch, $\bar{n}_{el}^{B}/\bar{n}_{el}^{T}=0.77$, modulation
period, $\lambda=11$, temperature, $t=0.01$, and the average
filling factor, $\bar{\nu}_{av}^{tot}=1.34$, to calculate the
electrostatic quantities. We see in Fig.\ref{fig:donor} that the
top layer is in phase with the modulation, whereas the bottom
layers phase is shifted by an amount of $\pi/2$, where the
modulation strength is chosen to be one percent. For the given
parameters both layers are compressible and screening is still
linear (see Fig.\ref{fig:mod} region (i)), i.e. the electrons can
redistribute according to the applied external potential. A
similar case for single layer geometries has been extensively
studied in Ref.[\onlinecite{Siddiki02:Oxford,Siddiki03:125315}]
and the linear dielectric function is given by
$\epsilon(q)=1+1/a_{0}q$ ($\sim 19$ for our parameters). This
linear screening approximation breaks down when the amplitude of
the screened potential becomes equal to the Fermi energy of the
unmodulated system\cite{Wulf88:162}. From this we would expect
that the breakdown amplitudes of the two layers should be directly
proportional to the density mismatch. In order to test this and
examine the non-linear screening regime we increase the strength
of the modulation monotonously and look at the variation of the
screened potential defined by, \be
var[V^{T,B}]=V^{T,B}(x=0)-V^{T,B}(x=0.3\mu m)]/\Omega_2.\ee Figure
\ref{fig:mod}a presents this variation, for top (solid line) and
bottom (dashed line) layers. In the regime denoted by (i) both
layers are compressible and the density profile can be
characterized similar to Fig.\ref{fig:donor}. With increasing the
modulation amplitude first the top layer enters to the non-linear
screening regime, since incompressible strips are formed (see
Fig.\ref{fig:mod}b). This is pronounced as a jump in the variation
an saturates at $\delta\sim 0.05$ (regime (ii)), meanwhile the
bottom layer is still compressible, (i.e in the linear regime) and
compensates (screens) the potential fluctuation generated [cf.
discussion related to Fig.\ref{fig:potfluc}b]. From the saturation
point we can easily predict the similar point of the bottom layer
to be $\delta\sim 0.07$. Actually we see that regime (iii) of
Fig.\ref{fig:mod}a starts at the expected value, where
incompressible strips are present at both layers showing a
characteristic distribution similar to Fig.\ref{fig:potfluc}c.
Here screening becomes linear again,
 but now with $\epsilon(q) = 1 + (D_T /D_0)/(a_0q)$,
where $D_T$ is the thermodynamic DOS in a Landau Level, which can
be estimated by $D_T /D_0 \sim \hbar\omega_c/(4k_BT)$. For
$\delta>0.15$ the bottom layer is split to five narrower channels
(see Fig.\ref{fig:mod}d) and the linear screening is broken around
$\delta\sim 0.16$. From the analytical expressions given in
Ref.[\onlinecite{Siddiki02:Oxford}] one can estimate the breakdown
amplitude as \be \label{eq:breakdown} \delta_{bd}\sim
\frac{2E_{F}^{B}.\epsilon(q)}{E_{0}}\sim 0.16, \ee which indeed is
in very good agreement with the obtained numerical result. The
small difference is due to the higher temperature and the decay of
mutual Coulomb interaction in growth direction, likewise the
difference at the slope of the plateaus. We also observe that the
plateaus occur at the integer multiples of the individual filling
factors, similar to the single layer system. Since we start with a
situation where the average filling factors of the both layers are
below two we see only one plateau, as discussed for the single
layer geometries the number of the plateaus depend on the average
filling factor without modulation. Finally we would like to note
that the linear screening will break down for the top layer for
$\delta\sim 0.2$ which can seen easily from Eq.\ref{eq:breakdown}.
\begin{figure}[h]
{\centering
  \includegraphics[width=.8\linewidth,angle=-90]{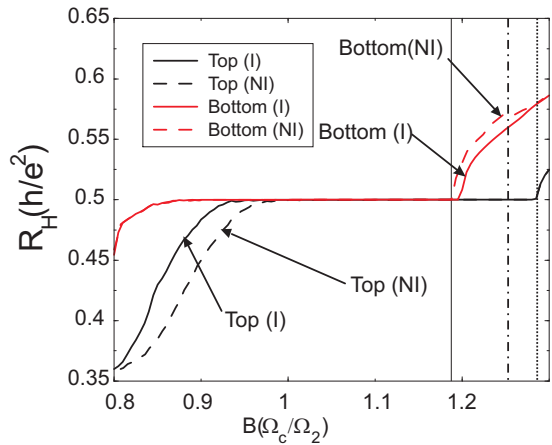}
\caption{ \label{fig:res} Comparison of the interacting
(solid-lines) and non-interacting (dashed-lines) Hall resistance
curves for fixed electron densities at default temperature. The
Hall resistance curve is shifted for the top layer as the
potential fluctuation created by the bottom layers incompressible
strip.}}
\end{figure}

The potential fluctuations and the donor modulation discussed
above plays an important role when one considers a fixed external
current flowing through both of the layers. These fluctuations are
generated by local charge imbalances, with respect to zero field
density distributions, and can be observed in the interval where
the other layer has an incompressible strip as they are poorly
screened.
\section{The Hall resistance curves\label{sec:hrc}} In this section we
investigate the effects of potential fluctuations on the Hall
resistances using the findings of SG\cite{siddiki2004} within the
linear response regime. In their work it was concluded that the
(long-range) potential fluctuations can widen, stabilize and shift
the quantized Hall (QH) plateaus as they affect the position and
the existence of the incompressible strips\cite{Siddiki:ijmp}. In
addition to the long-range fluctuations, in this section we also
include the potential fluctuations generated by the local charge
imbalances. Here we use the general expressions derived in
Ref.[\onlinecite{siddiki2004}], for a given electron density and
fixed current, to calculate the Hall and longitudinal resistances,
of the bilayer system. We also note that the averaging procedure
of the conductivities is carried over a length scale
($\lambda_{av}$), which is comparable with the Fermi wavelength
and, in particular is set to be $40$nm in our calculations. We
apply these results to two cases considered in
Sec.~\ref{sec:potentialfluctuations}, namely the interacting and
non-interacting systems. We remind that, in the case of
non-interacting system, the active layer does not have information
about the density inhomogeneities caused by the incompressible
strips of the passive layer. Hence comparing the resistance curves
of these two cases essentially gives a method to extract the
effect of the incompressible strips on the other layer.

In order to investigate the relation between the incompressible
strips of the passive layer and the magneto-transport coefficients
of the active layer qualitatively, we calculate the Hall
resistances for a magnetic field interval, where QH plateaus are
observed for both layers around filling factor two. In
Fig.~\ref{fig:res} we show the Hall resistances (in units of the
von Klitzing constant) {\sl vs.} magnetic field for interacting
(solid-lines) and non-interacting (dashed lines) systems. We start
the discussion with the high-magnetic field regime
($\Omega_c/\Omega_2>1.27$, i.e right side of the vertical dotted
line), which essentially corresponds to a density distribution
similar to that shown in Fig.~\ref{fig:potfluc}a. As there are no
incompressible strips within both layers, no noticeable potential
fluctuations are created due to local charge imbalance, therefore
the Hall resistances of both layers are the same for interacting
and non-interacting cases. If we examine the QH regime of the top
layer (the regime between vertical dash-dotted line and dotted
line in Fig.\ref{fig:res}), it is seen that at least one
incompressible strip is formed (cf. Fig.~\ref{fig:potfluc}b),
creating a potential fluctuation to the bottom layer. Meanwhile,
the bottom layer is still compressible all over the sample, so the
fluctuation can be screened nearly perfect, leading to a new
distribution of the electrons and a small change in the $R_{H}$
curves, due rearrangement of the local conductivities. The
situation is fairly different, in the interval
$1.18<\Omega_c/\Omega_2<1.25$ (the region between thin vertical
solid-line and dashed-dotted line), where incompressible strips
are formed at both of the layers (density distributions
corresponding to Fig.~\ref{fig:potfluc}c and
Fig.~\ref{fig:potfluc}d). In this regime, where both layers
produce fluctuations due to local charge imbalance and this
fluctuation can not be screened perfectly everywhere, which should
result in a difference in the Hall resistance curves. This is
observed for the bottom layer at the high-magnetic field edge,
since the perturbation slightly shifts the maximum magnetic field
value of the QH plateau. In this regime the quantized value of the
Hall resistance of the top layer does not change, since it only
depends on the presence of the incompressible strip. In the
$1<\Omega_c/\Omega_2<1.18$ interval of Fig.\ref{fig:res} there
exists two stable incompressible strips in both layers surrounded
by compressible regions, thus both layers are in the QH plateau.
Here we should note that, although the potential fluctuations are
created by the incompressible strips and are not screened
perfectly, this perturbation changes only the positions or the
widths of the incompressible strips, but not the value of the
$R_{H}$ within the plateau regime. Eventually it depends on the
amplitude of the perturbation, i.e. if the amplitude is large
enough to destroy the incompressible strips one does not observe
the quantized value for $R_{H}$. The change in $R_{H}$ can be
observed for the top layer at somewhat smaller values of the
magnetic field strength ($0.8\lesssim \Omega_{c}/\Omega_{2}
\lesssim1.0$) since the perturbation enlarges the plateau. In fact
the fluctuation widens the incompressible strips, with respect to
the non-interacting case and creates incompressible strips larger
than the averaging length, $\lambda_{av}$. Finally one ends with a
wider plateau. The difference in the Hall resistance curves
between the interacting and non-interacting systems, tends to
disappear as the incompressible strips become narrower and move
towards the edges by decreasing the magnetic field. This is
consistent with the previous observation, that the amplitude of
the fluctuation depends on the width of the incompressible region.

In summary  four magnetic field intervals are observed: (i) both
layers are compressible and there exists no difference for the
$R_{H}$ curves, calculated for the interacting and non-interacting
case, (ii) top layer, at least, has an incompressible strip and
creates a fluctuation, which shifts for the bottom layer the edge
of the Hall plateau to higher magnetic field values, (iii) both
layers show incompressible regions, the perturbation generated by
the bottom layer widens the incompressible strips of the top
layer, leading to a wider plateau, ({\sl iv}) the incompressible
strips of both layers become narrower and move towards the edges
and the fluctuation becomes inefficient, hence the difference is
smeared out. It is useful to mention that the \emph{interacting}
system is an \emph{equilibrium} solution, in the sense that the
electrons of both layers rearrange their distribution until a full
convergence is obtained (within a numerical accuracy), for a given
external potential profile.

\section{\label{sec:experiments}Comparison to the experiments}

Here we report on the magneto-transport hysteresis observed in the
bilayer systems measured at the Max-Planck Institut-Stuttgart by
S. Kraus\cite{siddikikraus:05}. In the experiments discussed here,
the magneto-resistances are measured as a function of the applied
perpendicular magnetic field. The sweep direction dependence is
investigated for matched and mismatch densities. Similar findings
have already been presented in the literature\cite{tutuc,pan},
however, have been discussed in a different theoretical content.

The samples are GaAs/AlGaAs double quantum well structures grown
by molecular beam epitaxy (MBE). Separate Ohmic contacts to the
two layers are realized by a selective depletion
technique\cite{Eisenstein90:2324}. A technique developed by Rubel
et al.[\onlinecite{Rubel98:207}] was used to fabricate backgates.
The metal gate on top of the sample acts as frontgate. The samples
were processed into $80\mu$m wide and $880\mu$m long Hall bars.
The high mobility samples are grown at the Walter-Schottky
Institut. They have as grown densities in the range $1.5-2.5
\times 10^{11}$cm$^{-2}$ and mobility is 100 m$^2$/Vs per layer.
The barrier thickness is 12nm and the quantum wells are 15nm wide.
The experiments were performed at low-temperatures (T » 270mK) and
the imposed current is always in the linear response regime
(I$\sim 50$nA). Other details of the experimental setup and
samples could be found in an upcoming
publication\cite{siddikikraus:05}.

\begin{figure}[h]
{\centering
  \includegraphics[width=.8\linewidth,angle=-90]{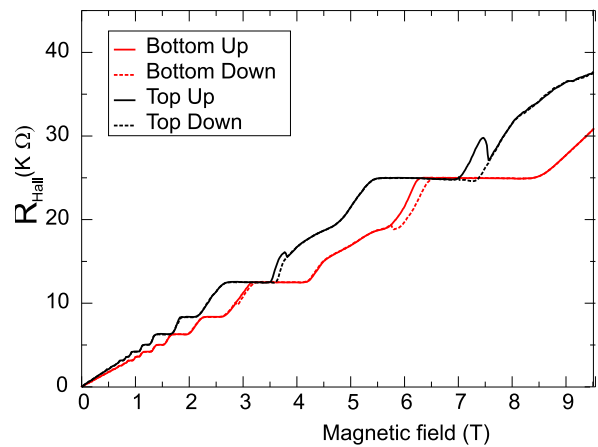}%
\caption{ \label{fig:experiment} The Hall resistances as a
function of magnetic field for a GaAs electron bilayer system with
a barrier width 12nm, considering two sweep directions. Data is
taken at a magnetic sweep rate $0.01$T/min. A strong hysteresis
develops in $R_{H}$ when the other layer is in a plateau. No
significant effect is observed at low field regime.}}
\end{figure}

We show the measured Hall resistances of the top layer (solid
lines for up and dashed lines for down sweep) and the bottom layer
(dotted lines for up and dashed-dotted lines for down sweep) as a
function of magnetic field strength and direction in
Fig.\ref{fig:experiment}. The data were taken at a sweep rate
$0.01$T/min and the base temperature is always kept at $270$mK.
Apparently it is seen that the resistances of the layers follow
different traces depending on whether the sweep is in up or down
direction, in certain $B$ field intervals. We see that the sweep
direction has no effect on the resistances at high magnetic fields
($B>8$T), where we would not expect\cite{siddiki2004} to have any
incompressible strips (see e.g. Fig\ref{fig:potfluc}a). The first
exiting feature is observed when the bottom layer is in the
plateau regime and the top layer is approaching to its plateau
regime ($6.5$T$<B<8.5$T). It is seen that the Hall resistance of
the top layer follows two different traces depending on the
sweeping direction, meanwhile the bottom layers resistance is
insensitive and assumes the quantized value.

The plateau regime is extended up to $B\sim 7.25$T for down sweep
and ends at $B=7$T for up sweep, meanwhile the bottom layer is
also in the QH regime. It is easy to understand why the QH regime
is wider for the down sweep considering the widths of the
incompressible strips as follows: while coming form high magnetic
field to low fields we first encounter with a wide incompressible
region at the bottom layer (see for example
Fig.\ref{fig:potfluc}b, of course here bottom layer has a higher
density) that generates a large potential fluctuation yielding a
wide plateau. However, for the other sweep direction two narrow
incompressible strips are observed first (Fig.\ref{fig:potfluc}d),
which creates relatively weak fluctuations, that have a negligible
effect on the plateau width. For smaller values of the magnetic
field, we observe at the top layer that up and down sweep curves
follow the same trace and have the quantized value at $B=6.8$T. A
similar hysteresis behavior is seen at the bottom layer resistance
curves ($5.7$T$<B<6.5$T), now the top layer is in the plateau
regime for both sweeping directions. The resistances at low
magnetic field interval ($4$T$<B<7$T) do not show a prominent
difference for two sweep directions. A less pronounced repetition
of the hysteresis is observed at the $\nu= 2$ plateau regimes,
e.g. $3.4$T$<B<3.6$T and $2.9$T$<B<3.2$T. A common feature (also
observed at different density mismatches) is that the hysteresis
is seen at the active layer only if the passive layer is in the
plateau regime. No hysteresis is observed if the QH regimes of the
layers coincide (see discussion in
Ref.[\onlinecite{siddikikraus:05}]). We summarize the experimental
observations as follows: the hysteresis is observed in the active
layer only if the passive layer is in the plateau regime.

We should mentioned that the hysteresis effect is a
non-equilibrium effect and "interacting" system is an equilibrium
solution, hence can not explain this effect directly. However, it
is easy to attribute the widening of the top layer's plateau (for
down sweep) to the potential fluctuations created by the bottom
layer, as already shown for the "interacting" case in
Fig.\ref{fig:res}. In accord with our numerical results (previous
section) obtained for the "non-interacting" system, the measured
quantities are independent of their history. We claim that the
measured resistances are the analogues of the calculated
resistances for the non-interacting system for the up sweep of the
top layer, since the incompressible strips are narrow and the
potential fluctuations are ineffective.

\section{Summary}
In this work we have studied the screening properties of an
electron-electron bilayer system. The electrostatic part is solved
numerically using a self-consistent screening theory by exploiting
the slow variation of the confining potential. We compared the
electron distributions for vanishing magnetic field and
temperature with the finite ones, and observed that a local charge
imbalance is created due to the formation of incompressible
strips. These dipolar strips produce external potential
fluctuations, as a function of applied magnetic field, to the
other layer. We have investigated properties of this potential
fluctuation by comparing the interacting and non-interacting
systems for a few characteristic $B$ values and obtained the Hall
resistances by a local scheme proposed in our recent
work\cite{siddiki2004}. We considered these fluctuations as a
perturbation to the other layer and observed that they widen,
stabilize and shift the plateaus as expected. In
Sec.\ref{sec:experiments} we have report and attempted to obtain
qualitative arguments for the hysteresis-like behavior by
reconsidering the symmetry breaking of the sweep direction, based
on the long relaxation times of the incompressible regions. This
is done by considering the "interacting" and "non-interacting"
schemes. Our results show that the Hall resistance curves follows
different paths if both of the layers have incompressible strips
within the sample. The amplitude of the deviation depends on the
widths and the positions of the incompressible regions.
\acknowledgements I would like to thank S. Kraus, S. C. J. Lok, W.
Dietsche and K. von Klitzing for their discussions and providing
experimental data. I am indebted to R. R. Gerhardts for many
valuable criticism and suggestions and also to S. Mikhailov for
initial discussions.


\end{document}